\documentclass[journal, draftcls, 			onecolumn]{IEEEtran}
\usepackage{cite}      
\usepackage{graphicx}  
\usepackage{psfrag}    
\usepackage{subfigure} 
\usepackage{multirow}
\usepackage{amsmath,amssymb,amsfonts}

\usepackage{booktabs}   

\usepackage{color}
\usepackage[normalem]{ulem} 

\usepackage[usenames,dvipsnames]{pstricks}
\usepackage{epsfig}
\usepackage{pst-grad} 
\usepackage{pst-plot} 

\graphicspath{{pix/}}

\newcommand{\matIN}[1]{#1}
\newcommand{\matOUT}[1]{}

\DeclareMathOperator{\D}{d\!}
\DeclareMathOperator{\sfT}{\mathsf T} 
\DeclareMathOperator{\sfH}{\mathsf H}
\newcommand{\nn}{\nonumber \\}
\newcommand{\mb}{\mathbf}
\newcommand{\Lt}{$L^2$}

\DeclareMathOperator{\RE}{Re}

\begin{document}

\title{Separable Implementation of L2-Orthogonal STC CPM with Fast Decoding}


\author{Matthias Hesse$\!~^1$, Jerome Lebrun$\!~^1$, Lutz Lampe$\!~^2$ and Luc Deneire$\!~^1$\vspace{1em}\\

\small
\begin{tabular}{cc}
 $~^1$Laboratoire I3S, CNRS & $~^2$Department of Electrical and Computer Engineering\\
 University of Nice, Sophia Antipolis & University of British Columbia\\
 Sophia Antipolis -- France & Vancouver -- Canada\\
 e-mail: hesse,lebrun,deneire@i3s.unice.fr & e-mail: lampe@ece.ubc.ca
\end{tabular}\\\normalsize\vspace{1em} \texttt{\small- final version - 28.Feb.2008 -}\vspace{-2.5em}}

\maketitle
\thispagestyle{empty}

\begin{abstract}
  ~In this paper we present an alternative separable implementation of
  $L^2$-orthogonal space-time codes (STC) for continuous phase modulation
  (CPM). In this approach, we split the STC CPM transmitter into a single conventional
  CPM modulator and a correction filter bank. While the CPM modulator
  is common to all transmit antennas, the correction filter bank
  applies different correction units to each antenna. Thereby 
  desirable code properties as orthogonality and full diversity are achievable 
  with just a slightly larger bandwidth demand.
  This new representation has three main advantages. First, it allows
  to easily generalize the orthogonality condition to any arbitrary
  number of transmit antennas. Second, for a quite general set of
  correction functions that we detail, it can be proved that full
  diversity is achieved. Third, by separating the modulation and
  correction steps inside the receiver, a simpler receiver can be
  designed as a bank of data independent inverse correction filters
  followed by a single CPM demodulator.
  Therefore, in this implementation, only one correlation filter bank for the
  detection of all transmitted signals is necessary. The decoding
  effort grows only linearly with the number of transmit antennas.
\end{abstract}


\section{Introduction}

The combination of space-time coding (STC) with continuous phase
modulation (CPM) systems has attracted considerable interest. It
brings the possibilities \matIN{of capacity increase \cite{Tela99}  and robustness to fading \cite{Taro98} in} systems that 
display good spectral and power efficiency \cite{Ande86}.
Pioneered by Zhang and Fitz \cite{Zhan00}, the first
STC CPM constructions were based on trellis codes. This approach was also
pursued by Zaji\'c and St\"uber in \cite{Zaji05} for full response
CPM, further optimized in \cite{Zaji06} and extended to partial
response CPM in \cite{Zaji07}. Bokolamulla and Aulin \cite{Boko07} and 
Maw and Taylor \cite{Maw05} designed STC by splitting the CPM signal in a memoryless modulator 
and a continuous phase encoder (CPE) \cite{Rimo88}. While Bokolamulla 
and Aulin use codes from \cite{Hamm00}, the latter combines an external encoder with the STC 
and the CPE. However, for these codes the decoding
effort grows exponentially with the number of transmit antennas.
This was partially circumvented by burst-wise orthogonality as
introduced by Silvester et al. in \cite{Silv05} and by
block-wise orthogonality as established by Wang and Xia in
\cite{Wang04}\cite{Wang05}.  Unfortunately, this latter design is based 
on the Alamouti code \cite{Alam98} and thus is restricted to two transmit antennas. 
An extension to 4 transmit antennas based on quasi orthogonal space-time codes 
was presented in \cite{Wang03}.

Mainly motivated by the low complexity of decoding as described in
\cite{Wang04}\cite{Wang05}, our present contribution concerns
orthogonal space-time block codes (STBC) for CPM systems.  In our previous
work \cite{Hess08a,Hess08b,Hess08c},  we have been
able to design $L^2$-orthogonal space-time codes for 2 and 3 transmit antennas which achieve full
rate and full diversity with low decoding effort. In \cite{Hess08a} we
generalized the two-antenna code proposed by Wang and Xia
\cite{Wang05} from pointwise to $L^2$-orthogonality. In
\cite{Hess08b} we presented the first $L^2$-orthogonal code family, coined
  {\it Parallel Codes} (PC), for CPM with 3 antennas. 

In the present paper, we briefly review some
of our previous results and generalize them to an arbitrary number of transmit antennas. 
More specifically, for {\it Parallel Codes} we
present an alternative approach to the encoding  by splitting
the STC CPM transmitter into a conventional CPM modulator and a correction filter
bank. While the modulator is shared by all transmit antennas, the correction filter bank is specific to each transmit antenna. Therefore, the correction filter bank fully characterizes the properties of
the code, e.g. orthogonality, diversity and coding gain.
This simple framework makes it possible to readily design 
$L^2$-orthogonal {\it Parallel Codes} for an arbitrary number of transmit
antennas and we prove that full diversity is achieved with these codes.

Again, by separating the demodulation and inverse correction steps at the
receiver side, a simple receiver is designed as a data 
independent inverse correction filter bank followed by a single decorrelation unit. In this
implementation, only one decorrelation unit for the detection of all
transmitted CPM signals is necessary. The overall decoding effort
grows only linearly with the number of transmit antennas. 

The remainder of the paper is organized as follows. In Section \ref{sec:code}, we present our new code representation, show that full diversity is achieved and give a condition to obtain $L^2$-orthogonality for an arbitrary number of transmit antennas. In Section \ref{sec:decode}, we introduce a fast decoding algorithm for {\it Parallel Codes}. In Section \ref{sec:sim}, the code performance and the decoding algorithm are evaluated by simulations and finally, in Section~\ref{sec:concl}, we conclude this paper.

\section{Generalized code representation}
\label{sec:code}

In this section we develop a simplified representation for $L^2$-orthogonal PC and prove that PC with linear phase correction functions provide full diversity. Finally, we give a condition to obtain $L^2$-orthogonal codes.

\subsection{System Model and Code Structure}
\label{sec:notation}

Let us briefly introduce our model for the CPM transmitter with $L_t$
transmitting antennas. We adopt the block structure from \cite{Hess08b} and accordingly we define the CPM signal for blocks of $L_t$ symbol intervals. The $l^{\mbox{th}}$ CPM block of length $L_tT$ is given by \cite{Ande86}
\begin{equation}
  s(t,\mb d)=\sqrt{\frac{E_B}{TL_t}}\exp\Big(j2\pi\Big[ \theta(l)+h\sum\limits_{i=lL_t-\gamma+2}^{(l+1)L_t}d_{i}q(t-iT) \Big]\Big).
  \label{eq:longS}
\end{equation}
for $lL_tT\leq t< (l+1)L_tT$. Here, $E_B$ is the block energy, $T$ is the symbol length, $\gamma$ the CPM memory length, $h=m_0/p$ is the modulation index with $m_0$ and $p$ relative primes and $d_i$ is the data symbol
taken from the set $$\Omega_d=\{ -M+1,-M+3, \ldots , M-3,M-1 \}.$$
For convenience, the data symbols of the current block $l$ are collected in the vector
$\mb d=\begin{bmatrix} d_{lL_t+1}&\hdots&d_{(l+1)L_t}\end{bmatrix}$. The phase pulse
$q(t)$ is a continuous function with $q(t)=0$ for $t\leq 0$ and
$q(t)=1/2$ for $t\geq\gamma T$ and the accumulated phase
\begin{equation}
  \theta(l)=\frac{h}{2}\sum\limits_{i=-\infty}^{lL_t-\gamma+1}d_i
\end{equation}
sums all $L_t$ symbols reaching $1/2$ till the end of the previous
block.

\begin{figure}
  \centering 
\scalebox{0.85} 
{
\begin{pspicture}(0,-2.4709375)(10.1171875,2.5109375)
\psframe[linewidth=0.02,dimen=outer](1.2782812,1.9290625)(0.17828126,1.3290626)
\psline[linewidth=0.02cm,arrowsize=0.05291667cm 2.0,arrowlength=1.4,arrowinset=0.4]{->}(1.2782812,1.6290625)(2.1782813,1.6290625)
\psframe[linewidth=0.02,dimen=outer](3.1782813,1.9290625)(2.1782813,1.3290626)
\usefont{T1}{ptm}{m}{n}
\rput(0.69640625,1.6390625){CPM}
\psframe[linewidth=0.02,dimen=outer](3.1782813,1.1290625)(2.1782813,0.5290625)
\psframe[linewidth=0.02,dimen=outer](3.1782813,-0.1709375)(2.1782813,-0.7709375)
\usefont{T1}{ptm}{m}{n}
\rput(2.6054688,1.6390625){$c_1(t)$}
\usefont{T1}{ptm}{m}{n}
\rput(2.675469,-0.4609375){$c_{L_t}(t)$}
\usefont{T1}{ptm}{m}{n}
\rput(2.6054688,0.8390625){$c_2(t)$}
\psline[linewidth=0.02cm](1.6782813,1.6290625)(1.68,0.6109375)
\psline[linewidth=0.02cm,arrowsize=0.05291667cm 2.0,arrowlength=1.4,arrowinset=0.4]{->}(1.6782813,0.8290625)(2.1782813,0.8290625)
\psline[linewidth=0.02cm,arrowsize=0.05291667cm 2.0,arrowlength=1.4,arrowinset=0.4]{->}(1.6782813,-0.4709375)(2.1782813,-0.4709375)
\psdots[dotsize=0.12](1.6782813,1.6290625)
\psdots[dotsize=0.12](1.6782813,0.8290625)
\psline[linewidth=0.02cm](3.1782813,1.6290625)(4.0782814,1.6290625)
\psline[linewidth=0.02cm](4.0782814,1.6290625)(4.0782814,2.1290624)
\psline[linewidth=0.02cm](4.0782814,1.8290626)(3.8782814,2.1290624)
\psline[linewidth=0.02cm](4.0782814,1.8290626)(4.278281,2.1290624)
\psline[linewidth=0.02cm](3.1782813,0.8290625)(4.0782814,0.8290625)
\psline[linewidth=0.02cm](4.0782814,0.8290625)(4.0782814,1.3290626)
\psline[linewidth=0.02cm](4.0782814,1.0290625)(3.8782814,1.3290626)
\psline[linewidth=0.02cm](4.0782814,1.0290625)(4.278281,1.3290626)
\psline[linewidth=0.02cm](3.1782813,-0.4709375)(4.0782814,-0.4709375)
\psline[linewidth=0.02cm](4.0782814,-0.4709375)(4.0782814,0.0290625)
\psline[linewidth=0.02cm](4.0782814,-0.2709375)(3.8782814,0.0290625)
\psline[linewidth=0.02cm](4.0782814,-0.2709375)(4.278281,0.0290625)
\psline[linewidth=0.02cm,arrowsize=0.05291667cm 2.0,arrowlength=1.4,arrowinset=0.4]{<-}(0.67828125,1.9290625)(0.67828125,2.3290625)
\usefont{T1}{ptm}{m}{n}
\rput(2.685469,0.2390625){$\vdots$}
\psline[linewidth=0.02cm,arrowsize=0.05291667cm 2.0,arrowlength=1.4,arrowinset=0.4]{->}(5.378281,1.6290625)(6.1782813,1.6290625)
\psline[linewidth=0.02cm](5.378281,1.6290625)(5.378281,2.1290624)
\psline[linewidth=0.02cm](5.378281,1.8290626)(5.1782813,2.1290624)
\psline[linewidth=0.02cm](5.378281,1.8290626)(5.5782814,2.1290624)
\psframe[linewidth=0.02,dimen=outer](7.1782813,1.9290625)(6.1782813,1.3290626)
\psframe[linewidth=0.02,dimen=outer](7.1782813,1.1290625)(6.1782813,0.5290625)
\psframe[linewidth=0.02,dimen=outer](7.1782813,-0.1709375)(6.1782813,-0.7709375)
\usefont{T1}{ptm}{m}{n}
\rput(6.605469,1.6390625){$c_1^*(t)$}
\usefont{T1}{ptm}{m}{n}
\rput(6.6754684,-0.4609375){$c_{L_t}^*(t)$}
\usefont{T1}{ptm}{m}{n}
\rput(6.605469,0.8390625){$c_2^*(t)$}
\usefont{T1}{ptm}{m}{n}
\rput(6.6854687,0.2390625){$\vdots$}
\psline[linewidth=0.02cm](5.778281,1.6290625)(5.778281,0.6290625)
\psline[linewidth=0.02cm,arrowsize=0.05291667cm 2.0,arrowlength=1.4,arrowinset=0.4]{->}(5.778281,0.8290625)(6.1782813,0.8290625)
\psline[linewidth=0.02cm,arrowsize=0.05291667cm 2.0,arrowlength=1.4,arrowinset=0.4]{->}(5.778281,-0.4709375)(6.1782813,-0.4709375)
\psdots[dotsize=0.12](5.778281,1.6290625)
\psdots[dotsize=0.12](5.778281,0.8290625)
\usefont{T1}{ptm}{m}{n}
\rput(0.39546874,2.2990625){$d_i$}
\psline[linewidth=0.02cm,linestyle=dotted,dotsep=0.16cm,arrowsize=0.05291667cm 2.0,arrowlength=1.4,arrowinset=0.4]{->}(4.278281,1.8290626)(5.0782814,1.8290626)
\psline[linewidth=0.02cm,linestyle=dotted,dotsep=0.16cm,arrowsize=0.05291667cm 2.0,arrowlength=1.4,arrowinset=0.4]{->}(4.378281,1.0290625)(5.0782814,1.7290626)
\psline[linewidth=0.02cm,linestyle=dotted,dotsep=0.16cm,arrowsize=0.05291667cm 2.0,arrowlength=1.4,arrowinset=0.4]{->}(4.278281,-0.2709375)(5.0782814,1.5290625)
\psframe[linewidth=0.02,dimen=outer](8.278281,1.9290625)(7.5782814,1.3290626)
\psframe[linewidth=0.02,dimen=outer](8.278281,1.1290625)(7.5782814,0.5290625)
\psframe[linewidth=0.02,dimen=outer](8.278281,-0.1709375)(7.5782814,-0.7709375)
\usefont{T1}{ptm}{m}{n}
\rput(7.9354687,1.6390625){$h_1^*$}
\usefont{T1}{ptm}{m}{n}
\rput(7.9054685,-0.4609375){$h_{L_t}^*$}
\usefont{T1}{ptm}{m}{n}
\rput(7.9354687,0.8390625){$h_2^*$}
\usefont{T1}{ptm}{m}{n}
\rput(7.8854685,0.2390625){$\vdots$}
\usefont{T1}{ptm}{m}{n}
\rput(3.6054688,1.0390625){$s_2(t)$}
\usefont{T1}{ptm}{m}{n}
\rput(3.605469,-0.2609375){$s_{L_t}(t)$}
\usefont{T1}{ptm}{m}{n}
\rput(3.6054688,1.8390625){$s_1(t)$}
\usefont{T1}{ptm}{m}{n}
\rput(4.5354686,2.0390625){$h_1$}
\usefont{T1}{ptm}{m}{n}
\rput(4.5354686,1.5390625){$h_2$}
\usefont{T1}{ptm}{m}{n}
\rput(4.3054686,0.4390625){$h_{L_t}$}
\psline[linewidth=0.02cm,arrowsize=0.05291667cm 2.0,arrowlength=1.4,arrowinset=0.4]{->}(7.1782813,1.6290625)(7.5782814,1.6290625)
\psline[linewidth=0.02cm,arrowsize=0.05291667cm 2.0,arrowlength=1.4,arrowinset=0.4]{->}(7.1782813,0.8290625)(7.5782814,0.8290625)
\psline[linewidth=0.02cm,arrowsize=0.05291667cm 2.0,arrowlength=1.4,arrowinset=0.4]{->}(7.1782813,-0.4709375)(7.5782814,-0.4709375)
\pscircle[linewidth=0.02,dimen=outer](9.278281,0.9290625){0.1}
\psline[linewidth=0.02cm](9.278281,1.0290625)(9.278281,0.8290625)
\psline[linewidth=0.02cm](9.178281,0.9290625)(9.378282,0.9290625)
\usefont{T1}{ptm}{m}{n}
\rput(5.7854686,1.8390625){$r(t)$}
\usefont{T1}{ptm}{m}{n}
\rput(8.725469,1.8390625){$x_1(t)$}
\usefont{T1}{ptm}{m}{n}
\rput(8.725469,1.1390625){$x_2(t)$}
\usefont{T1}{ptm}{m}{n}
\rput(8.765469,-0.2609375){$x_{L_t}(t)$}
\usefont{T1}{ptm}{m}{n}
\rput(9.8,-0.7){$x(t,d)$}
\psline[linewidth=0.02cm](8.278281,1.6290625)(9.278281,1.6290625)
\psline[linewidth=0.02cm,arrowsize=0.05291667cm 2.0,arrowlength=1.4,arrowinset=0.4]{->}(9.278281,1.6290625)(9.278281,1.0290625)
\pscircle[linewidth=0.02,dimen=outer](9.278281,-0.4709375){0.1}
\psline[linewidth=0.02cm](9.278281,-0.3709375)(9.278281,-0.5709375)
\psline[linewidth=0.02cm](9.178281,-0.4709375)(9.378282,-0.4709375)
\psline[linewidth=0.02cm,arrowsize=0.05291667cm 2.0,arrowlength=1.4,arrowinset=0.4]{->}(9.278281,-0.0709375)(9.278281,-0.3709375)
\psline[linewidth=0.02cm,arrowsize=0.05291667cm 2.0,arrowlength=1.4,arrowinset=0.4]{->}(8.278281,-0.4709375)(9.178281,-0.4709375)
\psline[linewidth=0.02cm](9.278281,0.8290625)(9.278281,0.6290625)
\psline[linewidth=0.02cm,linestyle=dotted,dotsep=0.16cm](9.278281,0.6290625)(9.278281,-0.0709375)
\psline[linewidth=0.02cm](5.778281,-0.0709375)(5.778281,-0.4709375)
\psline[linewidth=0.02cm,linestyle=dotted,dotsep=0.16cm](5.778281,0.6290625)(5.778281,-0.0709375)
\psline[linewidth=0.02cm,arrowsize=0.05291667cm 2.0,arrowlength=1.4,arrowinset=0.4]{->}(8.278281,0.9290625)(9.178281,0.9290625)
\psframe[linewidth=0.02,dimen=outer](9.778281,-0.9709375)(8.578281,-1.5709375)
\usefont{T1}{ptm}{m}{n}
\rput(9.195469,-1.2609375){$s^*(t,\tilde d)$}
\usefont{T1}{ptm}{m}{n}
\rput(1.7354688,1.8390625){$s(t,d)$}
\psline[linewidth=0.02cm,arrowsize=0.05291667cm 2.0,arrowlength=1.4,arrowinset=0.4]{->}(9.278281,-0.5709375)(9.278281,-0.9709375)
\psline[linewidth=0.02cm](9.278281,-1.5709375)(9.278281,-1.8709375)
\psframe[linewidth=0.02,dimen=outer](9.478281,-1.8709375)(9.078281,-2.4709375)
\psline[linewidth=0.02cm,arrowsize=0.05291667cm 2.0,arrowlength=1.4,arrowinset=0.4]{->}(9.078281,-2.1709375)(8.278281,-2.1709375)
\usefont{T1}{ptm}{m}{n}
\rput(8.4,-1.9609375){$D(d,\tilde d)$}
\usefont{T1}{ptm}{m}{n}
\rput(9.245469,-2.1809375){$\int$}
\psline[linewidth=0.02cm](1.68,-0.0890625)(1.68,-0.4690625)
\psline[linewidth=0.02cm,linestyle=dotted,dotsep=0.16cm](1.68,0.6109375)(1.68,-0.0890625)
\end{pspicture} 
}
  \caption{Block diagram of the transmitter and receiver for STC CPM
    using the generalized code representation}
  \label{fig:block}
\end{figure}
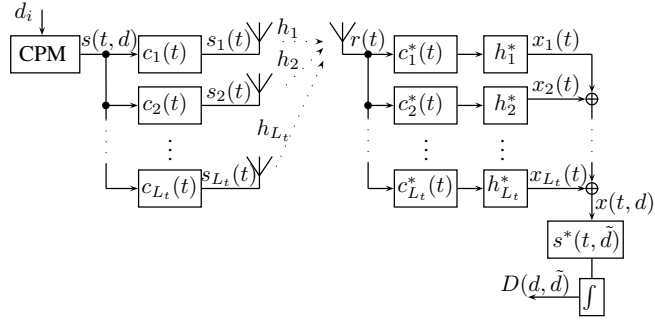

The family of $L^2$-orthogonal codes proposed in \cite{Hess08b} allows to send $L_t$ CPM signals over the transmit antennas. The signal sent by each antenna is further modified by an additional correction function. Here, we present a new, generalized representation for {\it Parallel Codes}, a member of the $L^2$-orthogonal code family. These codes use the same CPM signal $s(t,\mb d)$ for each antenna and only the correction function $c_m(t)$ differs for each antenna $m$. Consequently, we rewrite the vector of the transmitted signals $\mb s(t,\mb d)$ and obtain the new representation
\begin{equation}
  \mb s(t,\mb d)=s(t,\mb d)\mb c(t)=s(t,\mb d)\begin{bmatrix}c_1(t)\\\vdots\\c_{L_t}(t)\end{bmatrix}.
  \label{eq:defS}
\end{equation}
Figure \ref{fig:block} illustrates the single CPM modulator  and $L_t$ data independent correction functions for each transmitter antenna.  To maintain the constant amplitude
of the CPM signal, the correction functions modify only the phase, i.e.
\begin{equation}
  \label{eq:cmdef}
  c_m(t)=\exp\big(j2\pi\phi_{cm}(t)\big),
\end{equation}
where the design of $\phi_{cm}(t)$ will be described in the following.

\subsection{Diversity}
\label{sec:div}

For convenience, we assume a receiver equipped with only one antenna but
the extension to multiple antennas receivers is straightforward. The
channel between the $m^{\mathrm{th}}$ transmitting and the receiving antenna is
characterized by the channel coefficient $h_m$.  All channel
coefficients are assumed to be mutually independent, block-wise
constant, Rayleigh distributed random variables. Furthermore, we
assume perfect channel state information (CSI) at the
receiver and corruption by complex additive white Gaussian noise $n(t)$
(AWGN). Then the received signal follows as
\begin{align}
  r(t,\mb d)=&\mb h^{\sfT}\mb s(t,\mb d)+n(t)\\
  =&\begin{bmatrix} h_1&\hdots&h_{L_t}
  \end{bmatrix}
  \begin{bmatrix}
    c_1(t)\\\vdots\\c_{L_t}(t)
  \end{bmatrix}s(t,\mb d)+n(t).
\end{align}

To characterize STC with linear modulations, a signal matrix $\mb C_s$ was introduced in \cite{Taro98}. This matrix results from the correlation of all the possible differences of code words. To achieve full diversity, $\mb C_s$ ought to be full rank. It was shown by Zhang and Fitz \cite{Zhan03} that for nonlinear modulation, i.e. CPM here, the signal matrix should now be defined over waveforms, as
\begin{equation}
  \mb C_s=\int\limits_{0}^{L_lT}\mb\Delta(t)\mb\Delta^{\sfH}(t)\D t,
  \label{eq:Cs}
\end{equation}
where $\mb\Delta(t)$ is the difference between two transmitted signals
modulated by different data symbols $\mb d$ and $\mb{\tilde d}$
\begin{equation}
  \mb\Delta(t)=\begin{bmatrix}
    \Delta_1(t)\\
    \vdots\\
    \Delta_{L_t}(t)
  \end{bmatrix}=\mb s(t,\mb d)-\mb s(t,\mb{\tilde d}).
\end{equation}

Proposition 1 from \cite{Zhan03} shows that $\mb C_s$ has full rank
if and only if $\mb u^{\sfT}\mb\Delta(t)\neq0$ for all vectors $\mb
u\in\mathbb C^{L_t}$, except $\mb u=\mb 0$. This means that the waveforms of
the transmitted signals have to be linearly independent. By applying Eq.
(\ref{eq:defS}) we obtain the diversity condition
\begin{equation}
  \label{eq:conDiv}
  \mb u^{\sfT} \big(s(t,\mb d)-s(t,\mb {\tilde d})\big)\mb c(t)\neq0.
\end{equation}
Now, since $s(t,\mb d)$ and $s(t,\mb{\tilde d})$ are different for at least one
symbol, their difference is never zero for all $t$ within a block. Thus, Eq. (\ref{eq:conDiv}) simplifies to
\begin{equation}
  \label{eq:conDiv2}
  \sum\limits_{m=1}^{L_t}u_m c_m(t)=
  \sum\limits_{m=1}^{L_t}u_m \exp\big(j2\pi\phi_{cm}(t)\big) \neq0,
\end{equation}
which only depends on the correction function $c_m(t)$. A large class of functions fulfill Eq. (\ref{eq:conDiv2}). In the following, we focus only on  correction functions with linear phase. Thus we define  parametrized phase functions as
\begin{equation}
  \phi_{cm}(t)=\frac{m-1}{L_tT}\alpha t+\beta_m,
  \label{eq:phi}
\end{equation}
where $\beta_m$ is a constant phase offset and $\alpha$ is a nonzero slope. 
Now, $ \mb u^{\sfT} \mb c(t) = 0$ would imply that
\begin{equation}
  \label{eq:udel_null}
  \sum\limits_{m=1}^{L_t} \big(u_m \exp(j2\pi\beta_m)\big) \exp\left(j2\pi\frac{m-1}{L_tT}\alpha t\right)
  = 0 
\end{equation}
Introducing the polynomial
\begin{equation}
  p(x) = \sum\limits_{m=1}^{L_t}
  \big(u_m\exp(j2\pi\beta_m)\big)\, x^{m-1},
\end{equation}
Eq. (\ref{eq:udel_null}) would mean that $p(e^{j2\pi\alpha t/(L_t T)}) = 0 $.
This would imply that the polynomial $p(x)$, of degree $L_t-1$, vanishes
on more than $L_t$ different points. Thus, $p\equiv 0$ and $u_m = 0$
for all $m$.  Consequently, by \cite[Prop.~1]{Zhan03}, the signal
matrix $\mb C_s$ has full rank and all the codes achieve full
diversity. 

The linear phase correction functions are similar to the idea of tilting phase as proposed by Rimoldi \cite{Rimo88}. 
However, the purpose of tilted phase in \cite{Rimo88} was to simplify the states of single input single output CPM systems. Here, the phase drifts are introduced to achieve $L^2$-orthogonality between transmit antennas. Therefore the tilt angle (i.e. the slope of the linear phase function or the phase shift) has a quite different role in the two approaches.

\subsection{Orthogonality}
\label{sec:ortho}

With the new representation of CPM introduced in Section \ref{sec:notation} we
derive the orthogonality condition for an arbitrary number of transmit
antennas.  $L^2$-orthogonality is imposed by \cite{Hess08b}
\begin{align}
  E_B\mb I&=\int\limits_{0}^{L_tT}\mb s(t,\mb d)\mb s^{\sfH}(t,\mb d) \D t \nn
  &=\int\limits_{0}^{L_tT}\begin{bmatrix} c_1(t)\\\vdots\\c_{L_t}(t)
  \end{bmatrix}
  \begin{bmatrix}
    c_1^*(t)&\hdots&c_{L_t}^*(t)
  \end{bmatrix}
  \underbrace{s(t,\mb d)s^{*}(t,\mb d)}_{|s(t,\mb d)|^2=1} \D t \nn
  &=\int\limits_{0}^{L_tT}\begin{bmatrix}
    c_1(t)c_1^*(t)&\hdots&c_1(t)c_{L_t}^*(t)\\
    \vdots&&\vdots\\
    c_{L_t}(t)c_1^*(t)&\hdots&c_{L_t}(t)c_{L_t}^*(t)\\
  \end{bmatrix}\D t
\end{align}
where $\mb I$ is the $L_t\times L_t$ identity matrix. Due to the
constant amplitude of the CPM signal, orthogonality depends only on
the correction functions. By Def. (\ref{eq:cmdef}),
$c_m(t)c_{m}^*(t)=1$. So, we only need to cancel all the
crosscorrelation terms and get

\begin{align}
  0=&\int\limits_0^{L_tT}c_m(t)c_{m'}^*(t)\D t\\
  =&\int\limits_{0}^{L_tT}\exp\big(j2\pi[\phi_{cm}(t)-\phi_{cm'}(t)]\big)\D t\\
  =&\int\limits_{0}^{L_tT}\exp\left(j2\pi\left(\frac{m-m'}{L_tT}\alpha t
  +\beta_m-\beta_{m'}\right)\right)\D t \label{eq:ortho}
\end{align}
for $m\neq m'$. To fulfill Eq. (\ref{eq:ortho}) we have to integrate
over full rotations on the unit circle. Therefore, $\alpha$ needs to be
an integer. In the following we set $\alpha=1$ for two reasons:

\begin{enumerate}
	\item Minimizing bandwidth: The correction function causes a frequency shift depending on the slope of the phase. To minimize the overall bandwidth of the system the frequency shift needs to be small. Hence, the phase slope of the correction function is required to be minimal.
	\item Equivalence to {\it linPC} \cite{Hess08b}: If $\alpha=1$ {\it Parallel Codes} with linear phase function coincide with the {\it linPC} family proposed in \cite{Hess08b}. The phase offsets $\beta_m$ in Eq. (\ref{eq:phi}) correspond to the initial phases of the {\it linPC}. \matOUT{ which were investigated in \cite{Hess08c} for pure amplitude fading.} 
\end{enumerate}

\section{Fast Decoding Algorithm}
\label{sec:decode}

In this section we provide a simplified decoding scheme for the proposed parallel codes. For convenience, we assume only one receive antenna ($L_r=1$) but the extension to multiple antennas is straightforward.

\begin{figure}
  \centering \input{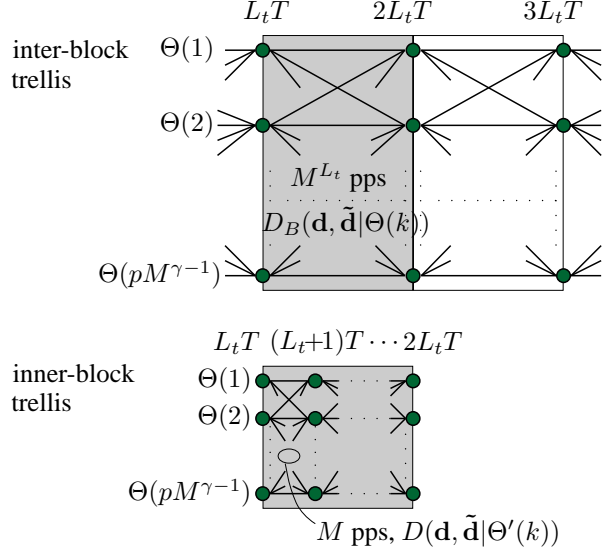}
  \caption{Merging of inter- and inner-block trellis for simplified
    detection with $l=1$ (pps - paths per state)}
  \label{fig:trellis}
\end{figure}

The received signal $r(t,\mb d)$ is a superposition of the transmitted CPM signals which are weighted by the channel coefficients. Due to the CPM inherent continuous phase encoder (CPE) \cite{Rimo88}, the received signal consists of $L_t$ superposing trellis codes. These are generally quite hard to decode. 

To reduce the complexity of the decoder we first consider the block structure of the proposed STC. This facilitates the splitting of the super trellis into an ``inter'' and ``inner-block'' trellis as shown in Figure \ref{fig:trellis}. To achieve full rate, each block contains $L_t$ symbols with an alphabet size $M$ which are distributed over the ST-block. Therewith each state of the inter-block trellis has $M^{L_t}$ leaving paths, i.e. in order to calculate all block distances $D_B(\mb d, \mb{\tilde d}|\Theta(k))$, $M^{L_t}$ matched filter of length $L_tT$ have to be applied $pM^{\gamma-1}$ times. This exponential growth of complexity with the number of transmit antennas makes the application in real world systems impossible. 

Eq. (\ref{eq:MLdec1}) shows the non-simplified maximum likelihood metric for the inter-block trellis. The absolute value contains all $L_t$ data symbols. Therefore we have to consider all crosscorrelations between the data symbols which gives us the previously mentioned $M^{L_t}$ path metrics. By using the $L^2$-orthogonality from the previous section those correlations are canceled out and we get
\begin{align}
 D_B(\mb d,\mb{\tilde d}|\Theta(k))=&\int\limits_{0}^{L_tT}\Big|r(t,\mb d)-\sum_{m=1}^{L_t}h_m^* s^*(t,\mb{\tilde d})c_m^*(t)\Big|^2\D t \label{eq:MLdec1}\\
  =&\sum_{m=1}^{L_t}\int\limits_{0}^{L_tT}\Big|r(t,\mb{\tilde d})-h_m s(t,\mb{\tilde d})c_m(t)\Big|^2\D t. \label{eq:MLdec2}
\end{align}

Here it can be seen that for nonlinear modulations, \Lt-orthogonality is sufficient to decorrelate the signals of the transmit antennas. 
\matOUT{ Orthogonal codes for linear modulations as in \cite{Taro98} are also sufficient to simplify Eq. (\ref{eq:MLdec1}). But they impose stronger restrictions upon the STC design.}
\matIN{The pointwise orthogonality of the orthogonal codes used in linear modulations is also a sufficient condition to simplify Eq. (\ref{eq:MLdec1}). But this would impose stronger restrictions upon the STC.}

Eq. (\ref{eq:MLdec2}) implies that $L_t$ conventional CPM signals have to be decoded. Hence, the complexity grows only linearly with the number of transmit antennas, i.e. due to decorrelation of the transmitted signals, each data stream from one transmit antenna can be decoded separately. Alternatively, the ML metric can be transformed into a correlation between the received signal $r(t,\mb d)$ and an hypothetical version of this signal. We get an equivalent correlation based metric by
\begin{align}
  D_B(\mb d,\mb{\tilde d}|\Theta(k))=&\sum_{m=1}^{L_t}\int\limits_{0}^{L_tT}\RE\Big\{r(t,\mb d)h_m^*c_m^*(t)s^*(t,\mb{\tilde d})\Big\}\D t
  \label{eq:longDist}.
\end{align}

By splitting the correction filter $c_m(t)$ from the conventional CPM signal $s(t,\mb d)$, we define a pseudo received signal as
\begin{equation}
  x(t,\mb d)=r(t,\mb d)\sum_{0}^{L_t}h_m^*c_m^*(t).
\end{equation}
This signal corresponds to a {\it single} preprocessed CPM signal which is decoded by
\begin{equation}
  D_B(\mb d,\mb{\tilde d}|\Theta(k))=\int\limits_{0}^{L_tT}\RE\Big\{x(t,\mb d)s(t,\mb{\tilde d})\Big\}\D t.
  \label{eq:correlBl}
\end{equation}
Hence, only one CPM signal has to be decoded and we obtain a single inner-block trellis which is shown at the bottom of Figure \ref{fig:trellis}. The metric to compute the symbol-wise distances at time slot $r$ is given by 
\begin{equation}
  D(\mb d,\mb{\tilde d}|\Theta'(k))=\int\limits_{(r-1)T}^{rT}\RE\Big\{x(t,\mb d)s(t,\mb{\tilde d})\Big\}\D t.
  \label{eq:correlSymb}
\end{equation}

This additional complexity reduction is accomplished due to the parallel structure of the proposed code. Finally, since $\alpha=1$,  the phase drift per block is always an integer ($0\ldots L_t-1$) and therewith $c_m(0) = \big(c_m(L_tT)\big)_{\text{mod}\, 1}$. Thus, the accumulated phase memory $\theta(l)$ at the beginning and the end of each block is defined over the same set of rational numbers, i.e. $\Omega_{\theta}=\{0,1/p,\ldots,(p-1)/p\}$. The states of the inner trellis at the end of one block $\Theta(k)$ and the beginning of the next $\Theta'(k)$ are consequently equal and inner and inter-block trellis can be merged to one trellis. The new block independent trellis is equivalent to the trellis of one conventional CPM signal. This is consistent with the model we use for the modulation where only one CPM signal is modulated and the signals for the different transmit antennas are created by the phase correction functions. One can look at the phase correction filters (phasers) of the transmitter, the physical channel and the dephasers of the receiver as a single input single output pseudo channel. This pseudo channel benefits from the full diversity introduced by the correction filters whereas at the transmitter and receiver only a conventional coder and decoder are necessary.

Finally, it should be noticed that the complexity of the proposed receiver can be further decreased. Namely, methods proposed in literature (\cite{Hube89}) can be additionally applied to the CPM decoder.

\section{Simulation Results}
\label{sec:sim}

In this section, we evaluate the proposed transceiver implementation and the performance of the code by means of simulations. For all our simulations, we use a linear phase pulse with a length of
$2T$ (2REC) given by $q(t)=t/4T$ for $0\leq t\leq 2 T$, $q(t)=0$ for $t\leq 0$ and $q(t)=1/2$ for $t\geq 2 T$.
An alphabet of size $M=4$ with $\Omega_d=\{-3,-1,1,3 \}$ is used. \matIN{Further, we assume  blockwise transmission with  block length $L_b=130$. The channel coefficients $h_i$ have Rayleigh distributed amplitude and uniformly distributed phase. They are assumed to be constant during one block length $L_bT$ and the receiver has perfect knowledge of those coefficients.}

\begin{figure}
  \centering
  \includegraphics[width=0.5\columnwidth]{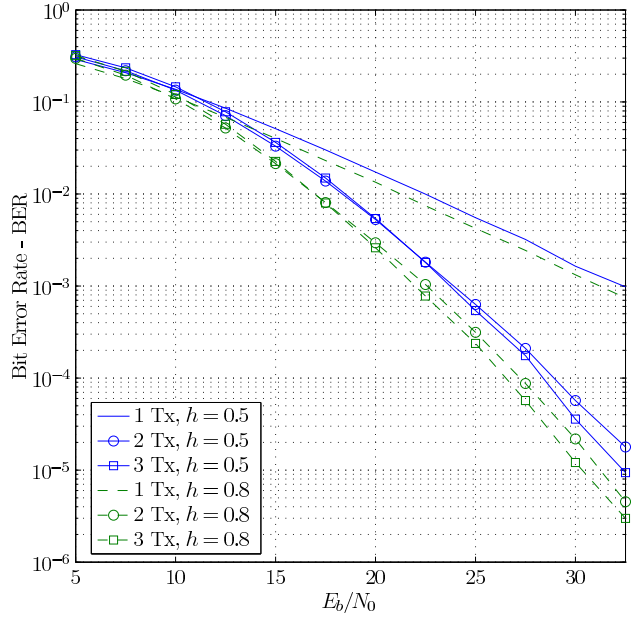}
  \caption{BER for PC CPM with $M=4$, 2REC, $h=0.5$ and $h=0.8$ }
  \label{fig:BER08}
\end{figure}

As stated earlier, the complexity of the
most costly part of the decoder, the MLSE, is independent of the
number of transmit antennas. In our case, the trellis has always
$pM=16$ states with $M=4$ paths originating from each state. \matIN{That means that we have to evaluate only $64$ path weights per symbol and $64 L_t$ per block. This is valid not only for one but also for three transmit
antennas.} In contrast, for $L_t=3$, a non-simplified receiver would have had to evaluate $pMM^{L_t}=1024$ paths per ST-block.  For the proposed scheme only the size of the correction filter bank grows with the number of transmit antennas. \matIN{Hence, the decoding effort grows only linearly with the number of transmitting antennas. Moreover, this filter bank needs to be evaluated only once per symbol. }

Figure \ref{fig:BER08} shows our simulation results for two different modulation indexes; $h=0.5$ and $h=0.8$. 
A larger modulation index increases the distance between two symbols and improves therewith the BER. The drawback of this improvement is an increased bandwidth. As expected, the simulations in Figure \ref{fig:BER08} show that the BER of the proposed STC CPM schemes also benefits from a larger modulation index. Further, the diversity gain becomes clearly visible. The slope of the BER curves increases with a growing number of transmit antennas.

For the second group of simulations ($h=0.8$) the decoding complexity
increases slightly due to the modified modulation index. The trellis
has now $pM=20$ states and we have to calculate $80$ path weights per
symbol.  The complexity of the correction filter bank remains
unchanged.

\section{Conclusion}
\label{sec:concl}

In this paper, we have presented a novel representation for $L^2$-orthogonal {\it Parallel Coded} CPM. This representation decouples the data-dependent CPM modulator from the antenna-dependent correction filter bank and enables the generalization of the $L^2$-orthogonal {\it Parallel Codes} to an arbitrary number of transmit antennas. It is also shown that these generalized codes achieve full diversity.

The main advantage of this representation arises at the
receiver level. The costly maximum likelihood sequence estimation,
necessary for decoding the CPM \cite{Hess08b}, is now implemented only once,
independently of the number of transmit antennas. The full diversity
of the system comes from the correction filter bank which is applied
only once per symbol. Hence, a simplified implementation and a
decoding effort that grows only linearly with the number of
transmit antennas is obtained in exchange for a slightly increased bandwidth for the correction filter.

\section*{Acknowledgment}
The work of M. Hesse is supported by a EU Marie-Curie Fellowship (EST-SIGNAL program) under contract No MEST-CT-2005-021175.

\bibliographystyle{IEEEtran}

\bibliography{IEEEabrv,bib}

\begin{thebibliography}{10}
\providecommand{\url}[1]{#1}
\csname url@samestyle\endcsname
\providecommand{\newblock}{\relax}
\providecommand{\bibinfo}[2]{#2}
\providecommand{\BIBentrySTDinterwordspacing}{\spaceskip=0pt\relax}
\providecommand{\BIBentryALTinterwordstretchfactor}{4}
\providecommand{\BIBentryALTinterwordspacing}{\spaceskip=\fontdimen2\font plus
\BIBentryALTinterwordstretchfactor\fontdimen3\font minus
  \fontdimen4\font\relax}
\providecommand{\BIBforeignlanguage}[2]{{%
\expandafter\ifx\csname l@#1\endcsname\relax
\typeout{** WARNING: IEEEtran.bst: No hyphenation pattern has been}%
\typeout{** loaded for the language `#1'. Using the pattern for}%
\typeout{** the default language instead.}%
\else
\language=\csname l@#1\endcsname
\fi
#2}}
\providecommand{\BIBdecl}{\relax}
\BIBdecl

\bibitem{Tela99}
I.~E. Telatar, ``Capacity of multi-antenna gaussian channels,'' \emph{European
  Trans. Telecommun.}, vol.~10, pp. 585 -- 595, 1999.

\bibitem{Taro98}
V.~Tarokh, N.~Seshadri, and A.~R. Calderbank, ``Space-time codes for high data
  rate wireless communication: Performance criterion and code construction,''
  \emph{IEEE Trans. Inf. Theory}, vol.~44, no.~2, pp. 744 -- 765, march 1998.

\bibitem{Ande86}
J.~Anderson, T.~Aulin, and C.-E. Sundberg, \emph{Digital Phase
  Modulation}.\hskip 1em plus 0.5em minus 0.4em\relax Plenum Press, 1986.

\bibitem{Zhan00}
X.~Zhang and M.~P. Fitz, ``Space-time coding for {R}ayleigh fading channels in
  {CPM} system,'' in \emph{Proc. of Annu. Allerton Conf. Communication,
  Control, and Computing}, 2000.

\bibitem{Zaji05}
A.~Zaji\'c and G.~St\"uber, ``Continuous phase modulated space-time codes,'' in
  \emph{Proc. of IEEE International Symposium on Communication Theory and
  Applications (ISCTA'05)}, July 2005, pp. 292 -- 297.

\bibitem{Zaji06}
------, ``Optimization of coding gain for full-response {CPM} space-time
  codes,'' in \emph{Proc. of IEEE Global Telecommunications Conference
  (GLOBECOM '06)}, Nov. 2006, pp. 1 -- 5.

\bibitem{Zaji07}
------, ``A space-time code design for partial-response {CPM}: Diversity order
  and coding gain,'' in \emph{Proc. of IEEE International Conference on
  Communications (ICC'07)}, June 2007, pp. 719 -- 724.

\bibitem{Boko07}
D.~Bokolamulla and T.~Aulin, ``Serially concatenated space-time coded
  continuous phase modulated signals,'' \emph{IEEE Tran. Wireless Commun.},
  vol.~6, no.~10, pp. 3487--3492, October 2007.

\bibitem{Maw05}
R.~L. Maw and D.~P. Taylor, ``Externally encoded space-time coded systems with
  continuous phase frequency shift keying,'' in \emph{Proc. Int. Conf. on
  Wireless Networks, Communications and Mobile Computing}, 2005, pp. 1597 --
  1602.

\bibitem{Rimo88}
B.~Rimoldi, ``A decomposition approach to {CPM},'' \emph{IEEE Trans. on Inf.
  Theory}, vol.~34, pp. 260 -- 270, March 1988.

\bibitem{Hamm00}
A.~R. Hammons and H.~E. Gamal, ``On the theory of space-time codes for {PSK}
  modulation,'' \emph{IEEE Trans. Inf. Theory}, vol.~46, pp. 524 -- 542, March
  2000.

\bibitem{Silv05}
A.~Silvester, R.~Schober, and L.~Lampe, ``Burst-based orthogonal {ST} block
  coding for {CPM},'' \emph{IEEE Trans. Wireless Commun.}, vol.~6, pp. 1208 --
  1212, April 2007.

\bibitem{Wang04}
G.~Wang and X.-G. Xia, ``An orthogonal space-time coded {CPM} system with fast
  decoding for two transmit antennas,'' \emph{IEEE Trans. Inf. Theory},
  vol.~50, no.~3, pp. 486 -- 493, March 2004.

\bibitem{Wang05}
D.~Wang, G.~Wang, and X.-G. Xia, ``An orthogonal space--time coded partial
  response {CPM} system with fast decoding for two transmit antennas,''
  \emph{IEEE Trans. Wireless Commun.}, vol.~4, no.~5, pp. 2410 -- 2422, Sept.
  2005.

\bibitem{Alam98}
S.~M. Alamouti, ``A simple transmit diversity technique for wireless
  communications,'' \emph{IEEE J. Sel. Areas Commun.}, vol.~16, no.~8, pp. 1451
  -- 1458, Oct. 1998.

\bibitem{Wang03}
G.~Wang, W.~Su, and X.-G. Xia, ``Orthogonal-like space-time coded {CPM} system
  with fast decoding for three and four transmit antennas,'' in \emph{Proc. of
  IEEE Global Telecommunications Conference (GLOBECOM '03)}, Nov. 2003, pp.
  3321 -- 3325.

\bibitem{Hess08a}
M.~Hesse, J.~Lebrun, and L.~Deneire, ``L2 orthogonal space time code for
  continuous phase modulation,'' in \emph{Proc. IEEE 9th Workshop on Signal
  Processing Advances in Wireless Communications SPAWC 2008}, 6--9 July 2008,
  pp. 401--405.

\bibitem{Hess08b}
------, ``Full rate {L2}-orthogonal space-time {CPM} for three antennas,'' in
  \emph{Proc. IEEE Global Telecommunications Conference IEEE GLOBECOM 2008},
  Nov. 30 2008--Dec. 4 2008, pp. 1--5.

\bibitem{Hess08c}
------, ``Optimized {L2}-orthogonal {STC} {CPM} for 3 antennas,'' in
  \emph{Proc. Wireless Communication Systems. 2008. ISWCS '08. IEEE
  International Symposium on}, 21--24 Oct. 2008, pp. 463--467.

\bibitem{Zhan03}
X.~Zhang and M.~P. Fitz, ``Space-time code design with continuous phase
  modulation,'' \emph{IEEE J. Sel. Areas Commun.}, vol.~21, pp. 783 -- 792,
  June 2003.

\bibitem{Hube89}
J.~Huber and W.~Liu, ``An alternative approach to reduced-complexity
  {CPM}-receivers,'' \emph{IEEE J. Sel. Areas Commun.}, vol.~7, no.~9, pp.
  1437--1449, Dec 1989.

\end{thebibliography}

\end{document}